\documentclass[prl,twocolumn]{revtex4}
\usepackage{graphicx}
\usepackage{hyperref}
\usepackage{amssymb}
\input epsf

\begin{document}

\title{Optical ferris wheel for ultracold atoms}

\author{\href{http://www.physics.gla.ac.uk/~sfrankea/}{S.\ Franke-Arnold},$^1$ J.\
Leach,$^1$ \href{http://www.physics.gla.ac.uk/Optics/Miles/}{M.\ J.\ Padgett},$^1$ V.\ E.\
Lembessis,$^2$ D.\,~Ellinas,$^3$ \href{http://www.photonics.ac.uk/staff/wright_a.htm}{A.\ J.\
Wright},$^4$ \href{http://www.photonics.ac.uk/staff/girkin_j.html}{J.~M.~Girkin},$^4$
\href{http://www.phy.hw.ac.uk/contacts/POhberg.htm}{P.~\"Ohberg},$^5$
\href{http://www.photonics.phys.strath.ac.uk/People/Aidan/Aidan.html}{A.~S.~Arnold}$^6$}
\affiliation{$^1$Dept.\ of Physics and Astronomy, SUPA, University of Glasgow, Glasgow G12 8QQ, UK\\
$^2$New York College, 38 Amalias Str., GR 105 58, Athens, Greece\\
$^3$Dept.\ of Sciences, Div.\ of Mathematics, Technical University of Crete, GR 731 00 Chania, Crete, Greece\\
$^4$Inst.\ of Photonics, SUPA, University of Strathclyde, Glasgow G4 0NW, UK\\
$^5$Dept.\ of Physics, SUPA, Heriot-Watt University, Edinburgh EH14 4AS, UK\\
$^6$Dept.\ of Physics, SUPA, University of Strathclyde, Glasgow G4 0NG, UK}
\date{\today}

\begin{abstract}
We propose a versatile optical ring lattice suitable for trapping cold and quantum degenerate atomic samples at discrete
angular positions.  We demonstrate the realisation of intensity patterns generated from Laguerre-Gauss ($\exp(i
\ell\theta)$) modes with different $\ell$ indices. The ring lattice can have either intensity maxima or minima, suitable
for trapping in red or blue detuned light, and it can be rotated by introducing a frequency shift between the Laguerre
Gauss modes.  The potential wells can be joined to form a uniform ring trap, making it ideal for studying persistent
currents and the Mott insulator transition in a ring geometry.
\end{abstract}

\maketitle


Confining ultracold atomic samples in optical lattices allows the investigation of effects
conventionally associated with condensed matter physics within a pure and controllable system.
Optical lattices have been employed to trap arrays of atoms \cite{Hemmerich} as well as Bose
condensates (BECs). Important experiments include the investigation of the quantum phase
transition from a superfluid to a Mott insulator \cite{Bloch}, and the realisation of arrays of
Josephson junctions \cite{Inguscio}.  Of particular interest is the study of quasi 1D systems as
quantum effects are strongest at low dimensionality.  An effective change of mass and associated
lensing have been observed in a moving 1D lattice \cite{Inguscio2}.  Various ring traps for
quantum degenerate gasses \cite{ram,gup} have been generated that are in many ways equivalent to
an infinite 1D geometry. More recently ring-shaped lattices have been proposed \cite{Amico}.

Optical beams at a frequency far detuned from the atomic or molecular resonance are one of the
fundamental tools for the manipulation of cold atoms and BECs \cite{Adams}.  The spatial structure
of the intensity distribution creates an energy potential well to trap and hold the target
species, either in the high intensity region of red detuned light, or in the low intensity region
of blue detuned light.  Translation of the intensity distribution of the beam can be used to
impart a global motion to the trapped atoms/molecules \cite{Meschede}.  Arbitrary intensity
patterns can be generated using spatial light modulators (SLMs) acting as reconfigurable
diffractive optical components, i.e.\ holograms.  Most notably SLMs have been employed to form
holographic optical tweezers \cite{Curtis} where a single laser beam is diffracted to form
multiple foci, trapping microscopic objects in complex 3D geometries \cite{Sinclair}. Very
recently, SLMs have also been used to manipulate single atoms \cite{Grangier} and BECs
\cite{Foot}. However, the nature of nematic liquid crystal devices means that most SLMs are
limited in their update rate to around $50\,$Hz, and even those based on ferroelectric
configurations are limited to $1\,$kHz \cite{Foot}. In this paper we establish a method for
creating both positive and negative optical potentials that can be rotated around the beam axis at
frequencies ranging from a few mHz to 100's of MHz -- optical ferris wheels for atoms or BECs. The
barriers between the individual potential wells can be controlled allowing the Mott transition
from a ring lattice to a uniform ring trap.

\vspace{-4mm}\section{Rotating ring lattice theory} \vspace{-4mm} Laguerre-Gauss (LG) beams have
an azimuthal phase dependence $\exp (i \ell \theta).$  The center of these beams contains a phase
singularity (optical vortex) where intensity vanishes.  By overlapping two co-propagating LG beams
with different $\ell$-values $\ell_1$ and $\ell_2=\ell_1+\delta\ell$, the beams interfere
constructively at $|\delta\ell|$ azimuthal positions, separated by regions of destructive
interference, leading to a transverse intensity profile comprising $|\delta\ell|$ bright or dark
petals. An angular frequency shift of $\delta\omega$ between the LG beams introduces an angular
petal rotation rate of $\delta\omega/ \delta\ell$ \cite{Courtial}.

\begin{figure}[!ht]
\vspace{-2mm}\includegraphics[width=\columnwidth]{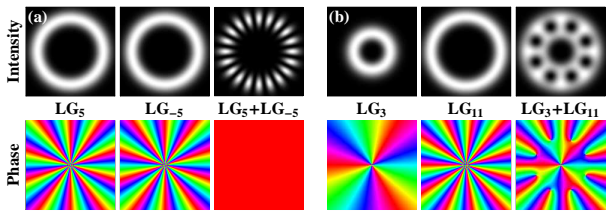}\vspace{-5mm}
\caption{\label{Fig:1}(color online) Generation of bright (a) and dark (b) lattices from
interfering LG beams with different $\ell$ values on an area of $6w\times6w$. Note that the dark
lattice sites are positioned at phase singularities.}
\end{figure}

Although LG beams with non-zero $p$-indices (i.e.\ with $p+1$ intensity rings), will allow more
freedom in the creation of exotic ring lattices, we confine our discussion in this paper to the
$p=0$ case as it already allows the simple, but highly adaptable, formation of both bright and
dark dynamic ring lattices.  We furthermore assume that the interfering LG beams have the same
focal position and beam waist $w_0$ in order to guarantee stable propagation. The scaled electric
field of an LG beam using a laser power $P$ at wavelength $\lambda$ can be expressed as:
\begin{equation}
{\rm LG}_{\ell} = A_{|\ell|}\exp\!\!\left[i\left(\!k(z-\frac{r^2}{2R})-\omega t+\Phi_{|\ell|}\!\!\right)\!\right]{\rm
e}^{-i \ell \theta}
\end{equation}
where $A_{|\ell|} = \sqrt{I} \sqrt{2/(\pi|\ell|!)}\left(\sqrt{2}r/w\right)^{|\ell|}
\exp(-r^2/w^2)$ is a dimensionless radial amplitude variation multiplied by the square root of a
beam intensity parameter $I=P w^{-2}$. Here $w=w_0\sqrt{1+(z/z_R)^2}$ is the beam waist, the
Rayleigh range is $z_R=\pi {w_0}^2/\lambda$, the radius of curvature is $R=z(1+(z_R/z)^2)$, and
$\Phi_{|\ell|}=(|\ell|+1)\arctan(z/z_R)$ the Gouy phase.  By interfering two LG beams with
different $\ell$ and angular frequency we obtain the intensity distribution:
\begin{eqnarray} \label{interf}
\!\!\!\!\!\!\!\!\!\!I&=&|{\rm LG}_{\ell_1}(\omega)+{\rm LG}_{\ell_2}(\omega+\delta\omega)|^2
\\&=&{A_{|\ell_1|}}^2\!+\!{A_{|\ell_2|}}^2\!+\!2 A_{|\ell_1|}A_{|\ell_2|}\cos\!\left(\delta\ell\,\theta-\delta\omega\,t+\delta\Phi\right) \nonumber
.
\end{eqnarray}
We have omitted the term $\frac{\delta\omega}{c}(z-\frac{r^2}{2R})$ in the cosine as it is
negligible for our experimental parameters. The Gouy phase difference
$\delta\Phi_{\delta\ell}=(|\ell_1|-|\ell_2|)\arctan(\frac{z}{z_R})$ can be significant near the
focus. One ring lattice site will rotate to the angle of the next site in a distance $\Delta
z=\tan(\frac{2\pi}{||\ell_1|-|\ell_2||})z_R$ from the focus, i.e.\ $\Delta z<z_R$ for
$||\ell_1|-|\ell_2||>8$. In our experiment we operate away from the focus so that the twist due to
the Gouy phase is negligible.
The spatial intensity in Eq.~\ref{interf} has $|\delta\ell|$ intensity maxima and minima as a function of $\theta$ and
rotates at an angular frequency $\delta\omega/\delta\ell$.  Complete constructive or destructive interference occurs at a
radius where both beams have equal intensity, determined by $A_{|\ell|}$.  For the case of $\ell_1=-\ell_2$ the
cylindrically symmetric intensity pattern comprises $2\ell$ petals (Fig.~\ref{Fig:1}(a)) \cite{Harris}, forming a bright
lattice. If $|\ell_1|\neq |\ell_2|$, the radii of the intensity rings differ. By choosing appropriate pairs of $\ell_1$
and $\ell_2$ one can generate dark lattices (Fig.~\ref{Fig:1}(b)).

The maximum intensity of a single LG$_\ell$ beam can be approximated to $I_\ell/(4\sqrt{|\ell|})$
at a radius $r_{\ell}\approx w\sqrt{|\ell|/2},$ \cite{Turnbull} and this approximation improves
for large $\ell$. One can also show that the electric field in the radial direction has a
full-width-half-maximum (FWHM) of $\sqrt{2 \ln(2)}w$. By choosing
$r_{\ell_2}-r_{\ell_1}\approx\sqrt{2 \ln(2)}w,$ (i.e.\ $\ell_2\in\mathbb{Z}$ with
$\ell_2\approx\pm (\sqrt{|\ell_1|}+2\sqrt{\ln(2)})^2)$, and $I_{\ell_2}=
\sqrt{|\ell_2/\ell_1|}I_{\ell_1},$ the two LG electric fields have similar maximum amplitudes and
are separated by 1 FWHM. This leads to a dark lattice with an approximately uniform depth in the
radial and azimuthal directions (Fig.~\ref{Fig:1}(b)). We also note that the intensity gradient
becomes maximal $\approx\sqrt{3}I_\ell/(4 w \sqrt{|\ell|})$ at $r\approx r_\ell\pm w/\sqrt{8},$
which can be used for determining lattice site stability at high rotation rates.

 \vspace{-4mm}
 \section{Rotating ring lattice experiment}
 \vspace{-4mm}
Precise laser frequency shifts can be produced by passing light through an acousto-optic modulator (AOM).  An acoustic
modulation of angular frequency $\omega_{\rm RF}$ applied to a crystal produces a traveling Bragg grating, shifting the
frequency of the first order diffracted beam by $\omega_{\rm RF}$.  Typically operating at around $\omega_{\rm
RF}/(2\pi)\approx 100\,$MHz, such modulators can be tuned over $10$'s of MHz. Two AOMs operating at $\omega_{{\rm RF}_1}$
and $\omega_{{\rm RF}_2}$ can produce light beams differing in angular frequency by $\omega_{{\rm RF}_1}-\omega_{{\rm
RF}_2}$ which can range from 0 to 10's of MHz. Our radio frequency signal generators (Marconi 2019) are passively highly
stable, but to ensure long term relative stability we synchronize their $10\,$MHz clocks. In order to eliminate the slight
angular shift produced by tuning the modulator frequency, the experiment is configured in a double-pass arrangement, thus
doubling the frequency shift to $\delta\omega=2(\omega_{{\rm RF}_1}-\omega_{{\rm RF}_2})$. We note that alternatively, a
small frequency shift can be imposed onto a light beam by passing circularly polarized light through a rotating half wave
plate \cite{Garetz}, which due to an accumulated geometric or Berry phase \cite{Simon}, shifts the frequency by twice the
rotation speed of the waveplate. This approach has been employed in optical tweezers \cite{Paterson}.

A Gaussian laser beam can be readily converted into a Laguerre-Gaussian mode by diffraction from a forked grating where
the positive and negative first order beams correspond to opposite signs of $\ell$ \cite{Bazhenov}.  In our experiment the
forked gratings are generated on a computer addressed SLM (HoloEye).  The mode purity of the diffracted Laguerre-Gaussian
beams is enhanced beyond standard hologram design by incorporating a spatially dependent modulation of the hologram
blazing \cite{Leach}.

\begin{figure}[b]
\includegraphics[width=1\columnwidth]{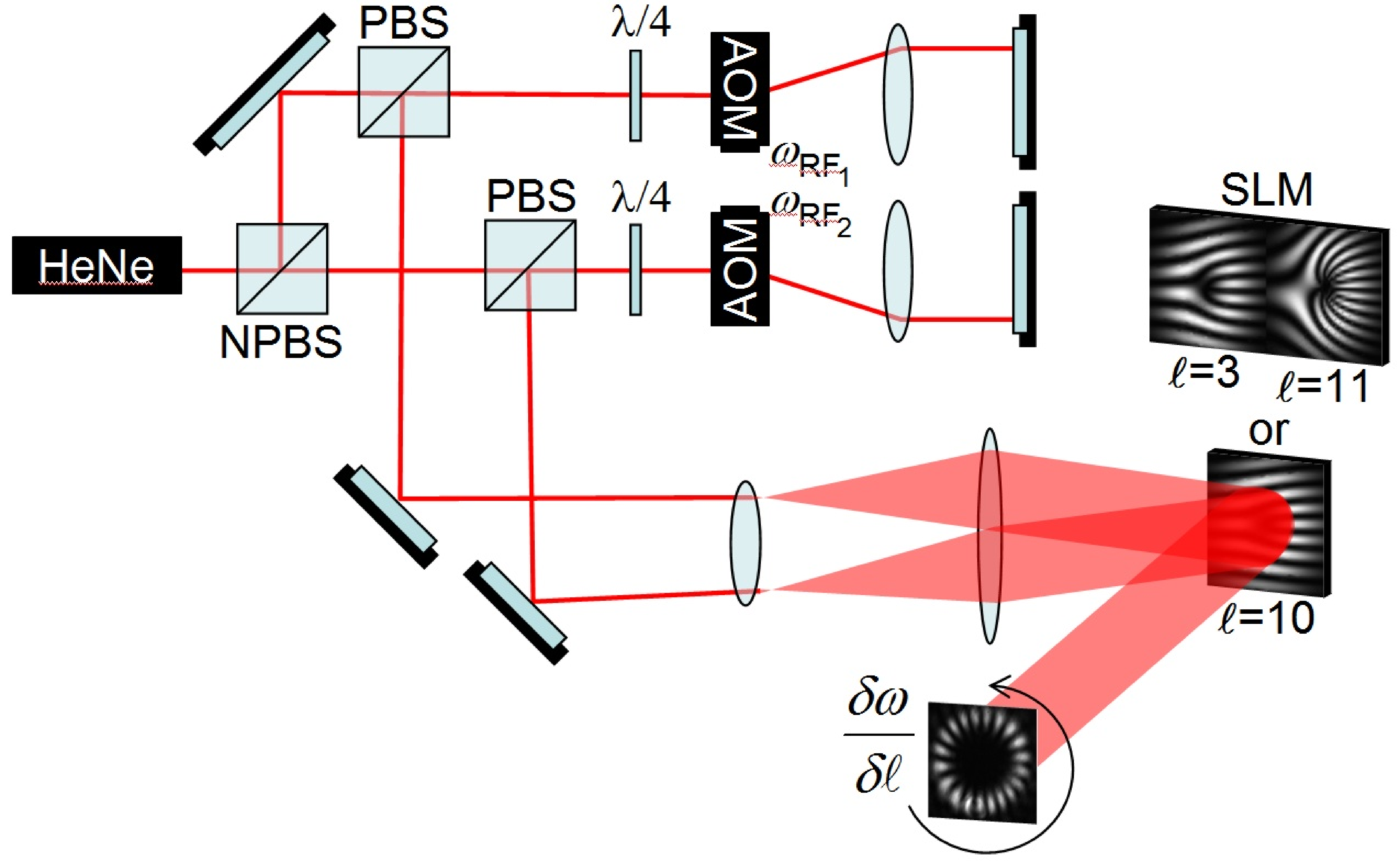}\vspace{-1mm}
\caption{\label{Fig:2}(color online) Experimental setup for generating rotating dark or bright
optical ring lattices.  Two double-passed AOMs impose a frequency shift between the light beams.
Bright lattices are generated by interfering the positive and negative diffracted beam from an
$\ell$ forked hologram, whereas dark lattices are obtained from two separate holograms.}
\end{figure}

Figure~\ref{Fig:2} shows the experimental arrangement used to create bright and dark rotating ring
lattices.
 The Gaussian beam from a helium-neon laser is divided and double passed through two
AOMs, leading to laser beams with an angular frequency difference of $\delta\omega$. These beams
are expanded to the size of the SLM.  For the bright lattice, the SLM is programmed with an
$\ell$-forked diffraction grating and the two beams are aligned such that the positive and
negative diffracted first-order, which have opposite signs of $\ell$, subsequently interfere to
give an intensity pattern rotating at angular frequency $\delta\omega/(2\ell)$.

For the dark lattice we need to overlap two appropriate Laguerre beams with order $\ell_1$ and
$\ell_2$. In our experiment we generated the required $\ell_1$ and $\ell_2$ forked holograms on
different parts of the same SLM, with each laser beam incident on one of the areas and aligned so
that the reflected beams are recombined to form the $|\ell_1-\ell_2|$ petalled dark lattice. We
note that alignment of the $\ell_1$ and $\ell_2$ beams is comparatively uncritical as the true
zero intensity at the dark lattice sites results from optical vortices (a $2\pi$ electric field
phase winding around the dark lattice site).

\begin{figure}[ht]
\vspace{-2mm}\includegraphics[width=1\columnwidth]{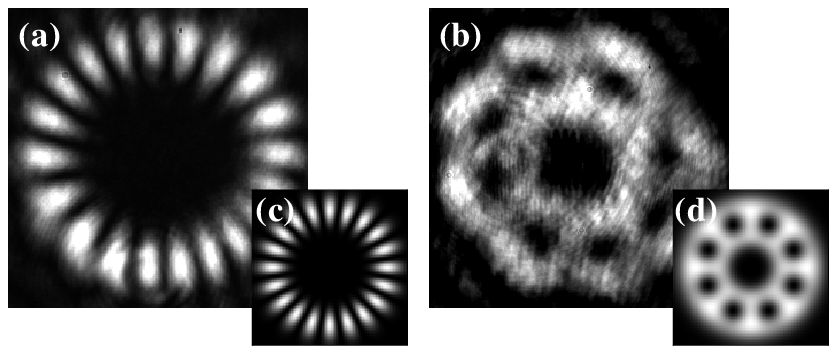}\vspace{-2mm}
\caption{\label{Fig:3}(color online) Observed intensity distribution for the bright (a) and dark
(b) lattice on an area of $3\times3\,$mm$^2$ and the corresponding theoretical distributions (c)
and (d).  The bright lattice is generated from LG beams $\ell_1=-\ell_2=10$ of equal intensity and
the dark lattice from $\ell_1=3,\ell_2=11$ with $I_2\approx \sqrt{\ell_2/\ell_1}I_1.$ As an
illustration of a rotating lattice we have made movies of the experiments e.g.\ (link
\href{http://www.physics.gla.ac.uk/Optics/Miles/lpm10.gif}{$\ell_1=-\ell_2=10$}).}
\end{figure}

Visualization of a rotating lattice requires high speed photography. Using shutter speeds down to
$5\,$ns we have observed the rotating intensity patterns for frequency shifts of up to $10\,$'s of
MHz between the two interfering Laguerre-Gaussian modes. The petal patterns rotate at the expected
frequencies. In Fig.\ \ref{Fig:3}(a) and (b) we show still images of the light and dark lattice
respectively, which agree well with theory.

 \vspace{-4mm}
 \section{Applications to atom optics}
 \vspace{-4mm}

By subjecting cold atoms to the dark or bright ring lattice described above, they can be trapped
in the resulting light potential.  In order to limit losses due to photon scattering we assume a
light beam far detuned from the atomic resonance. The AC Stark potential $U$, and photon
scattering rate $S,$ are related to the light intensity $I$, and detuning $\Delta=\omega-\omega_0$
by:
\begin{equation}
U\approx\frac{\hbar \Gamma^2 I}{8\Delta I_{\rm S}}, \;\;\;\;\;\;\;\;\;S\approx\frac{\Gamma^3 I}{8
I_{\rm S} \Delta^2},
\end{equation}
where $\Gamma$ and $I_{\rm S}$ denote the linewidth and saturation intensity of the atomic
transition, respectively. To illustrate the experimental feasibility of our scheme we use the
two-level dipole potential approximation, this could be extended to a higher-order multi-level
atom model \cite{approx}. We now consider the specific example of the D2 transition of $^{87}$Rb
atoms with $\Gamma=2\pi\times 6 {\rm MHz},$ $\lambda=780{\rm nm},$ $I_{\rm
S}=16.3\,$W$\,$m$^{-2}$. We assume a ring lattice laser total power of $2\,$W, which is focussed
to a beam waist of $w_0=30\,\mu$m at $1064\,$nm for trapping in the bright lattice and $660\,$nm
for trapping in the dark lattice. For a ring lattice with 10 potential wells this results in a
peak intensity of $5\times10^8\,$W$\,$m$^2$ corresponding to a potential well $65\,\mu$K deep for
the bright ($\ell_1\!=\!5\!=\!-\ell_2$) lattice and $0.8\times10^8\,$W$\,$m$^2$ corresponding to
$15\,\mu$K for the dark $(\ell_1=5,\ell_2=15)$ lattices respectively. The coldest atoms trapped in
the high intensity regions of the red detuned light potential will scatter a photon every $2\,$s.
For the blue detuned lattice the coldest atoms are trapped at dark lattice sites and scattering
will be negligible -- even the hottest atoms only scatter a photon every $6\,$s.

The optical lattice potential is sufficient to provide confinement in the transverse direction. To additionally localise
atoms in the axial $(z)$ direction we suggest a hybrid trap, combining the optical lattice with a quadrupole magnetic trap
\cite{kettoptplug,ram}. For the red lattice one could consider all-optical confinement in a tightly focused lattice with a
short Rayleigh range, but there is a trade-off between axial confinement and scattering rate. Instead, atoms could be
optically pumped into magnetic weak-field-seeking states and loaded into a quadrupole magnetic potential
$\textbf{B}=B_1\{x/2,y/2,-z\}$. The centre of the quadrupole field could be positioned away from the beam focus to ensure
a stable Gouy phase. However, for a standard quadrupole gradient of $B_1=100\,$G/cm, the atoms will be confined axially to
a region much smaller than the Rayleigh range and the twist of the Gouy phase becomes negligible. In this hybrid magnetic
and optical trap one can use standard RF evaporation, allowing in-situ cooling to quantum degeneracy. Circularly polarised
LG lattice beams are required to maintain the symmetry between the quadrupole magnetic field and the light field and
obtain a uniform ring lattice potential.

Alternatively, one can provide axial confinement in a ring lattice by using counterpropagating
laser beams to create a standing wave, generating an axially separated stack of $\delta\ell$
lattices similar to the method suggested in \cite{Amico}. However, by introducing a frequency
shift between the forward and backwards LG beam, the individual ring lattices will not only rotate
but also translate along the $z$-axis at a speed $\Delta \omega \lambda/(4 \pi)$. Additionally,
having a single ring lattice rather than a stack of ring lattices simplifies the experiment and
enables single-site addressability.

\begin{figure}[!h]
\vspace{-4mm}
\includegraphics[width=.87\columnwidth]{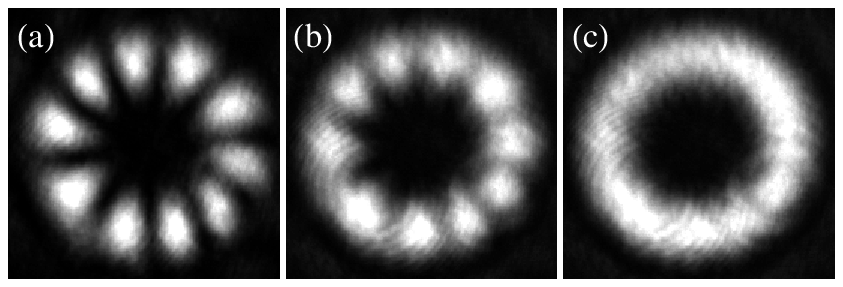}
\vspace{-5mm}
\includegraphics[width=.87\columnwidth]{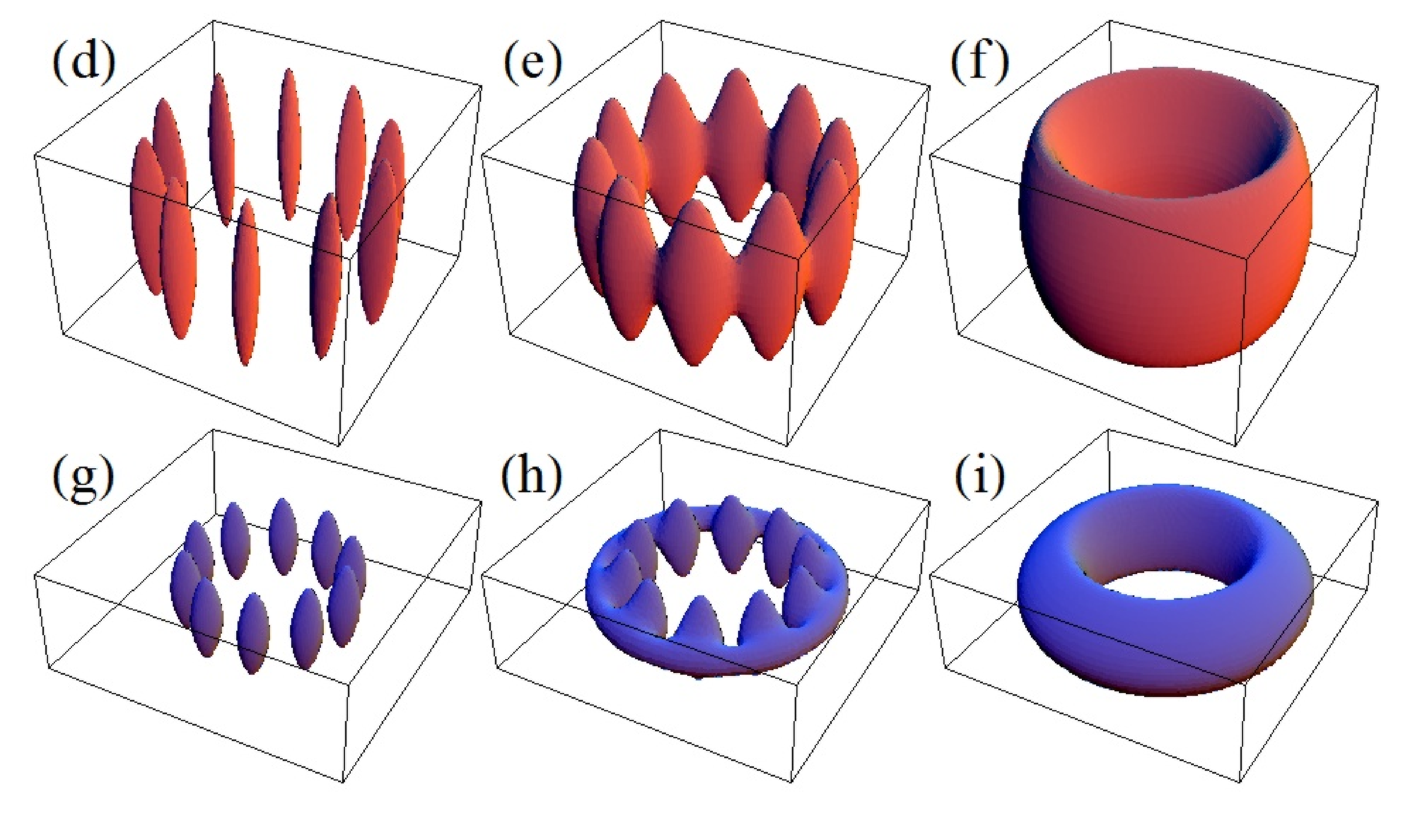}
\vspace{1mm} \caption{(color online) Lattices suitable for studying the Mott transition between a
10-site ring lattice and a ring trap. Images (a)-(c) are from optical experiments. Images (d)-(f)
((g)-(i)) depict a red (blue) detuned hybrid magnetic/optical lattice with
$\eta_1=1-\eta_2=0.5,0.99,1$ $(0.5,0.8,1)$ respectively. The red (blue) lattice contours are at
$15\,\mu$K ($12\,\mu$K), and the boxes have $xyz$ dimensions $120\times120\times80\,\mu$m$^3$
($260\times260\times80\,\mu$m$^3$).} \label{Fig:4}
\end{figure}

Our hybrid ring lattice enables the observation of the Mott insulator transition in a geometry
with periodic boundary conditions.  To adjust the barrier depth, and hence the tunneling between
sites, the relative power $\eta_{1,2}$ in the $\ell_{1,2}$ beams can be varied. Experimentally,
this can easily be achieved by varying the modulation amplitude of both AOMs while keeping the
overall light intensity constant.  To make full use of all laser power, an electro-optic modulator
could be used to rotate the polarisation from the laser incident on a polarising beamsplitter
leading to the two AOMs. For the bright lattice $\eta_{1,2}$ variation directly converts a uniform
ring into a ring lattice.  Images from our optical experiment are shown in Fig.~\ref{Fig:4}(a)-(c)
and the corresponding hybrid lattice theory in Fig.~\ref{Fig:4}(d)-(f). For the dark lattice, the
transition between uniform and multi-petalled ring is achieved by gradually dimming the outer LG
beam, and outer transverse confinement is then provided by the magnetic potential
(Fig.~\ref{Fig:4}(g)-(i)).

The dynamic nature of our lattice could also be used to initiate persistent currents. In order to
trap atoms in a rotating well pattern, several conditions need to be fulfilled: their initial
temperature must be low enough in order to be trapped, the rotation speed must change slowly
enough so that the atoms can adiabatically follow, and the centrifugal acceleration must be small
enough for the radial potential gradient. This constraint is much higher than the critical
rotation rate for vortex creation in 1D $\omega_c = \frac{\hbar}{4mR^2}\approx0.1\,$rad/s for our
parameters.

\section{Conclusions}\vspace{-5mm} We have experimentally obtained both bright and
dark optical ring lattices, with tunable barriers between sites, and with a tunable rotation rate.
Furthermore we have shown that, in combination with a magnetic trap, these lattices will be ideal
for studying quantum degenerate gases. Future applications of the lattice include studies of:
persistent currents, rotation of a ``quantum register,'' collisional studies using two
counter-propagating rings.

\textbf{Acknowledgements:} This work is supported by the UK EPSRC, and SFA is a Dorothy Hodgkin
Research Fellow of the Royal Society. VEL and DE were supported by `Pythagoras II' of the EPEAEK
programme and VEL was also supported by the CATS programme of the ESF (grant 756).

\end{document}